# THERMAL CONDUCTIVITY OF SEGMENTED NANOWIRES

Denis L. Nika*, Aleksandr I. Cocemasov

*E. Pokatilov Laboratory of Physics and Engineering of Nanomaterials, Department of Physics and Engineering, Moldova State University, Chisinau, MD-2009, Republic of Moldova*

Alexander A. Balandin*

*Department of Electrical and Computer Engineering and Materials Science and Engineering Program, University of California – Riverside, Riverside, California, 92521 USA*

## ABSTRACT

We present a review of the phonon thermal conductivity of segmented nanowires focusing on theoretical results for Si and Si/Ge structures with the constant and periodically modulated cross-sections. We describe the use of the face-centered cubic cell and Born-von Karman models of the lattice vibrations for calculating the phonon energy spectra in the segmented nanowires. Modification of the phonon spectrum in such nanostructures results in strong reduction of the phonon thermal conductivity and suppression of heat transfer due to a trapping of phonon modes in nanowire segments. Possible practical applications of segmented nanowires in thermoelectric energy generation are also discussed.

*Corresponding authors: D.L. Nika: dlnika@yahoo.com and A.A. Balandin: balandin@ece.ucr.edu

# I. INTRODUCTION

Phonons, i.e. quanta of the crystal lattice vibrations, affect all physical processes in solids [1]. They scatter the electrons and limit the electron mobility near room temperature (RT) and above and influence on the optical properties of crystalline materials. Acoustic phonons are the main heat carriers in insulators and semiconductors. In the long-wavelength limit, acoustic phonons in bulk crystals have nearly linear dispersion, which can be written as $\omega = V_S q$, where $V_S$ is the sound velocity, while the optical phonons are nearly dispersion-less and have a small group velocity $V_G = d\omega/dq$.

Spatial confinement of acoustic phonons in nanostructures affects their dispersion [2-8]. It modifies acoustic phonon properties such as phonon group velocity, polarization, density of states, and changes the way acoustic phonons interact with other phonons, defects and electrons [2-8]. Modification of phonon energy spectra and phonon scattering at the boundaries of nanostructures make the thermal conductivity of nanostructures much lower than the corresponding bulk values [2-8]. It has been demonstrated both experimentally and theoretically that the room temperature (RT) lattice thermal conductivity in freestanding Si nanolayers and nanowires is two orders of magnitude lower than the corresponding bulk value of Si [8-12].

Reduction of size and a corresponding increase in the density of elements on a chip make the task of improving thermal conductivity an important problem in modern electronics [5, 8, 13-15]. An insufficient heat removal from devices implies their overheating, which degrades their performance and limits the operating frequency. On the other hand, decrease of the thermal conductivity may improve the quality of thermoelectrics. The efficiency of the thermoelectric devices is determined by the figure of merit $ZT = S^2 \sigma T / (\kappa_{ph} + \kappa_{el})$, where $S$ is the Seebeck coefficient, $\sigma$ is the electrical conductivity, $T$ is the absolute temperature, $\kappa_{ph}$ and $\kappa_{el}$ is the lattice (phonon) and electronic thermal conductivity, respectively. One possible way to increase the figure of merit consists in the achieving low values of thermal conductivity when maintaining good electronic transport characteristics (electrical conductivity and Seebeck coefficient) in phonon-blocking and electron-transmitting nanostructures [5, 8, 16-18].

In spite of the fact that bulk Si is a poor thermoelectric with room-temperature *ZT* ~ 0.01 [19], thin Si, Ge and Si/Ge nanowires (NWs) are considered promising for thermoelectric applications [20-23] owing to the much lower values of the RT lattice thermal conductivity $\kappa_{ph}$ <1 W m$^{-1}$ K$^{-1}$. It has been demonstrated experimentally that the good electrical conductivity, as in doped bulk Si, and poor thermal conductivity of Si NWs [20] and rough Si NWs [21] provide relatively high values of *ZT* ~ 0.3 to 0.6 at RT. These experimental results stimulate theoretical search of one-dimensional Si-based nanostructured materials with the ultra-low thermal conductivity. The reduction of the RT lattice thermal conductivity up to 75% was theoretically demonstrated in Si/Ge core-shell NWs with Ge thicknesses of several monoatomic layers [24-25]. The corresponding enhancement of *ZT* in these NWs was also predicted [26].

In this chapter we review phonon and thermal properties of another kind of semiconductor nanostructures, promising for thermoelectric and thermo-isolating application due to their ultra-low thermal conductivity – referred as *segmented nanowires (SNW)*, i.e. nanowires consisting of segments of different size and/or made of different materials [27-31]. Following our recent publications [27-29] we describe how the thermal flux in segmented Si and Si/Ge nanowires can be suppressed by almost three orders-of-magnitude in comparison with that in bulk Si, and by an order-of-magnitude in comparison with generic Si nanowires. Fabrication of the cross-section modulated SNWs with layer thickness from several monoatomic layers (ML) up to several nanometers is still a technological challenge. However, reports of fabrication of InP, InN/InGaN or metallic SNWs [32-36] suggest that nanostructures considered in this chapter are feasible.

The rest of the chapter is organized as follows. In Section II we describe the theoretical models for phonons and discuss phonon properties of Si and Si/Ge segmented nanowires. In Section III we describe the method of calculation of the thermal conductivity in segmented nanowires and review their thermal properties. We give our conclusions in Section IV.

## II. PHONONS IN SEGMENTED NANOWIRES

The schemes of a generic rectangular nanowire and considered segmented nanowires are shown in Fig. 1: generic NW with the lateral cross-section $d_x \times d_y$ (panel a), Si/Ge SNW with segments dimensions $d_x \times d_y \times l_z^1$ and $d_x \times d_y \times l_z^2$ (panel b), cross-section modulated Si SNW (MSNW) with segments dimensions $d_x^1 \times d_y^1 \times l_z^1$ and $d_x^2 \times d_y^2 \times l_z^2$ (panel c) and cross-section modulated

Si/Ge SNW with silicon segments $d_x^1 \times d_y^1 \times l_z^1$ and $d_x^2 \times d_y^2 \times l_z^2$ covered by the Ge shell with thickness $d_{Ge}$ (panel d).

<Figure 1>

The external surfaces of the considered nanostructures are assumed to be free [37-39]. The *X* and *Y* axes of the Cartesian coordinate system are located in the plane of the cross-section of the nanowire and are parallel to its sides, while the axis *Z* is directed along the axis of the nanowire. We suppose that the length of the nanowire along the axis *Z* is infinite. The origin of coordinates is at the center of the cross-section of the nanowire. The translation period of SNW/MSNW consists of two segments with dimensions $d_x^1 \times d_y^1 \times l_z^1$ and $d_x^2 \times d_y^2 \times l_z^2$, respectively. The length period of SNW/MSNW is $L = l_z^1 + l_z^2$. The dimensions of nanowires considered in this chapter is presented in Table 1.

<Table 1>

## II.1. Face-centered cubic cell model of lattice dynamics in bulk crystals

The crystal lattice of silicon or germanium is a diamond-like lattice and it consists of two face-centered cubic sublattices, which are shifted along the main diagonal of a unit cell by 1/4 of its length. In the Face-centered cubic (FCC) model two shifted FCC sublattices are considered as a common FCC lattice with the double mass at each lattice node. This simplification neglects the optical phonon modes, but allows expressing three force constants of the model through independent elastic constants of a material. The latter allows one to simulate the acoustic properties of heterostructures consisting of layers with different acoustic properties and various dimensions [40-41].

In the FCC model all lattice nodes in a bulk crystal are translationally equivalent. The displacement of a lattice atom in the node labeled with the number $\vec{n}$, can be written as

$$\vec{u}(\vec{n};\vec{q},t) = \vec{w}(\vec{q})e^{i(\vec{q}\vec{n}-\omega t)}, \qquad (1)$$

where $\vec{w}(\vec{q})$ is the time-independent displacement. Because of the equivalence of nodes, the amplitude of their displacements does not depend on the node number: $\vec{n} = \vec{a}_1 n_1 + \vec{a}_2 n_2 + \vec{a}_3 n_3$, where $\vec{a}_1 = \frac{a}{2}(0,1,1)$, $\vec{a}_2 = \frac{a}{2}(1,1,0)$, $\vec{a}_3 = \frac{a}{2}(1,0,1)$ are basis vectors of the face-centered cubic cell, $n_1$, $n_2$ and $n_3$ are integers, and *a* is the lattice constant: $a$(Si)=0.549 nm, $a$(Ge)=0.565 nm. The node displacement in a bulk crystal is described by the equations of motion:

$$m\ddot{u}_i(\vec{n},\vec{q}) = F_i(\vec{n},\vec{q}), \qquad i = x, y, z. \qquad (2)$$

where $F_i(\vec{n},\vec{q})$ is a component of the force acting on the node $\vec{n}$ from the other nodes of the lattice and $m$ is the node mass (double atomic mass in the framework of FCC model). In the harmonic approximation

$$F_i(\vec{n},\vec{q}) = -\frac{\partial V}{\partial u_i(\vec{n},\vec{q})} = -\sum_{\vec{n}',j}\Phi_{ij}(\vec{n},\vec{n}')u_j(\vec{n}',\vec{q}), \qquad (3)$$

where $\Phi_{ij}(\vec{n},\vec{n}')$ is the three-dimensional matrix of the force constants and $V$ is the potential energy of the lattice.

Taking into account Eq. (1), one can obtain

$$m\omega^2 w_i(\vec{q}) = \sum_{j=1,2,3,\vec{h}} D_{ij}(\vec{q};\vec{h})w_j(\vec{q}), \qquad (4)$$

where $D_{ij}(\vec{q};\vec{h}) = \Phi_{ij}(0,\vec{h})e^{i\vec{q}\vec{h}}$ is the dynamic matrix and $\vec{h} = \vec{n}' - \vec{n}$.

We take into account the interaction of the node with the nearest and second-nearest nodes. The interaction with the 12 nearest nodes is centrally symmetric and it is described by one constant $\alpha_1^{FCC}$ [42]. The matrix of the force constants in this case is $\Phi_{il}(\vec{n}',\vec{n}) = -\alpha_1^{FCC}(\vec{n}',\vec{n})h_i^1 h_l^1/(h^1)^2$, where $\vec{h}^1$ indicates the positions of the nearest nodes of the node $\vec{n}=0$, $h_i^1$ is the projection of the vector $\vec{h}^1$ on the corresponding coordinate axis $X_i$. The interaction with the second-nearest nodes is not centrally symmetric and it is described by two constants, $\alpha^{FCC}$ and $\beta^{FCC}$ [43-44]. The vector $\vec{h}^2$ describes the position of the 6 second-nearest nodes of the node $\vec{n}=0$:

$$\begin{aligned}\Phi_{ij}(0,\vec{h}^2 = a(\pm 1,0,0)) &= \delta_{ij}\gamma_{ii}, \gamma_{11} = \alpha^{FCC}; \gamma_{22} = \gamma_{33} = \beta^{FCC} \\ \Phi_{ij}(0,\vec{h}^2 = a(0,\pm 1,0)) &= \delta_{ij}\gamma_{ii}, \gamma_{11} = \gamma_{33} = \beta^{FCC}; \gamma_{22} = \alpha^{FCC} \qquad (5)\\ \Phi_{ij}(0,\vec{h}^2 = a(0,0,\pm 1)) &= \delta_{ij}\gamma_{ii}, \gamma_{11} = \gamma_{22} = \beta^{FCC}; \gamma_{33} = \alpha^{FCC}.\end{aligned}$$

Because for every node with the vector $\vec{h}$ there exists a node with the vector $-\vec{h}$, the dynamic matrix is real. Comparing the phonon dispersions $\omega(q)$ for three phonon branches (one longitudinal and two transversal) obtained from Eq. 4 in the long-wave limit $q \to 0$ with those derived within a continuum approach [39], we established the following relations between the constants $\alpha_1^{FCC}, \alpha^{FCC}, \beta^{FCC}$ and the elastic moduli of a cubic crystal $c_{11}$, $c_{12}$ and $c_{13}$: $\alpha_1^{FCC} = a(c_{12}+c_{44})/2$, $\alpha^{FCC} = a(c_{11}-c_{12}-c_{44})/4$, $\beta^{FCC} = a(c_{44}-c_{12})/8$ [27].

**II.2. Born-von Karman model of lattice dynamics in bulk crystals**

The real unit cell structure is taken into account in Born-von Karman (BvK) model of lattice dynamics. For convenience, let's identify the atoms of the first sublattice of diamond − like lattice as the "dark" atoms while the atoms of the second sublattice as the "white" atoms as depicted in Figure 2.

<Figure 2>

The dynamic matrix in BvK model has the form: $D_{ij}(\vec{r}_k, \vec{r}'_k) = \Phi_{ij}(\vec{r}_k, \vec{r}'_k)/\sqrt{m(\vec{r}_k)m(\vec{r}'_k)}$, where $m(\vec{r}_k)$ [$m(\vec{r}'_k)$] is the mass of the atom at $\vec{r}_k$ [$\vec{r}'_k$], $\Phi_{ij}(\vec{r}_k, \vec{r}'_k)$ is the matrix of force constants and $\vec{h} = \vec{r}_k - \vec{r}'_k$. For the atom at $\vec{r}_k$, the summation in Eq. (4) is performed over all the nearest and second-nearest atoms at $\vec{r}'_k$. In the case of silicon or germanium, the atom at $\vec{r}_k$ has 4 nearest neighbors at $\vec{r}'_{k,n} = \vec{r}_k + \vec{h}_n^I$ (n=1,...,4) and 12 second-nearest neighbors at $\vec{r}'_{k,n} = \vec{r}_k + \vec{h}_n^{II}$ (n=1,..,12). The components of vectors $\vec{h}_n^I$ and $\vec{h}_n^{II}$ are presented in Table 2. In our BvK model, the interaction of an atom with its nearest and second-nearest neighbors is described by the following force constant matrices: $\Phi_{ij}^I = (16/a^2)(\alpha\delta_{ij} + \beta(1-\delta_{ij}))h_{n,i}^I h_{n,j}^I$ for the nearest atoms (n=1,...,4) and $\Phi_{ij}^{II} = (4/a^2)(\lambda\delta_{ij}(a^2/4 - h_{n,i}^{II} h_{n,i}^{II}) + \mu\delta_{ij} h_{n,i}^{II} h_{n,i}^{II} + \nu(1-\delta_{ij})h_{n,i}^{II} h_{n,j}^{II})$ for the second-nearest atoms (n = 1, ..., 12), where $\alpha$, $\beta$, $\mu$, $\lambda$ and $\nu$ are the force constants of BvK model, $\delta_{ij}$ is the Kronecker's delta and i,j=x,y,z. The force constant matrix $\Phi_{ij}(\vec{r}_k, \vec{r}'_k = \vec{r}_k)$ is obtained from the condition that the total force acting on the atom $\vec{r}_k$ at the equilibrium position is equal to 0, i.e. $\Phi_{ij}(\vec{r}_k, \vec{r}'_k = \vec{r}_k) + \sum_{\vec{r}'_k} \Phi_{ij}(\vec{r}_k, \vec{r}'_k \neq \vec{r}_k) = 0$.

By solving the equations of motion (4) at $\Gamma$ and $X$ Brillouin zone points of bulk Si or Ge, we expressed three constants $\alpha$, $\mu$, and $\lambda$ of our model through $\beta$ and the frequencies of the LO and TO phonons at $\Gamma$ point and the LA phonon at $X$ point:

$$\alpha = m\omega_{LO}^2(\Gamma)/8,$$
$$\mu = m\left(2\omega_{LA}^2(X) - \omega_{LO}^2(\Gamma)\right)/32, \quad (6)$$
$$\lambda = m\left(4\omega_{TO}^2(X) - 2\omega_{LA}^2(X) - \omega_{LO}^2(\Gamma)\right)/32 - \beta/2.$$

The constants $\beta$ and $\nu$ were treated as fitting parameters and were obtained from the best fit to experimental dispersion curves for bulk Si [45] and Ge [46]. The numerical values of the force constants for Si and Ge are indicated in the last column of Table 2.

<Table 2>

**II.3. Lattice dynamics in segmented nanowires**

In segmented nanowires, the displacements of the atoms (or nodes of FCC model) belonging to one period are independent, therefore the displacement amplitude $\vec{w}$ depends on the atomic coordinates. The rest of the atomic displacements are equivalent to those in the selected period due to the translational symmetry along the Z-axis. In the case of a generic NW, the translation period consists of two atomic layers of the "dark" atoms and two atomic layers of the "white" atoms (all layers are perpendicular to the Z axis). For SNW/MSNW, the number of atomic layers in the period is determined by L. The displacements of equivalent atoms have the form:

$$\vec{u}(x, y, z + n \cdot L; q_z, t) = \vec{w}(x, y, z; q_z)e^{i(q_z nL - \omega t)}, \qquad (7)$$

where $\vec{w}(x, y, z; q_z) \equiv \vec{w}(\vec{r}; q_z)$ is the displacement amplitude of the atom with coordinates $x$, $y$ and $z$; the period is labeled by an integer $n$, and $q_z$ is the phonon wavenumber. The equations of motion for the displacement are

$$\omega^2 w_i(\vec{r}_k; q_z) = \sum_{j=x,y,z; \vec{r}_k'} D_{ij}(\vec{r}_k, \vec{r}_k') w_j(\vec{r}_k'; q_z), \ k=1,\ldots,N, \ i = x,y,z, \qquad (8)$$

where $N$ is the number of atoms in the NW or SNW/MSNW translational period. To calculate the energy spectra of phonons in NWs and SNWs/MSNWs we numerically solve the set of equations (8) with free boundary conditions in the XY-plane, i.e. we assume that all force constants outside of the nanowire are equal to 0. The calculations were performed for all $q_z$ values in the interval $(0, \pi/a)$ for NWs and $(0, \pi/L)$ for SNWs/MSNWs. Invariance of the set (8) with respect to the reflection in the planes of symmetry leads to the following four possible types of solutions [47]:

Dilatational (D): $w_1^{AS}(x_1, x_2); w_2^{SA}(x_1, x_2); w_3^{SS}(x_1, x_2) \rightarrow w_i^D$;

Flexural$_1$ (Flex$_1$): $w_1^{AA}(x_1, x_2); w_2^{SS}(x_1, x_2); w_3^{SA}(x_1, x_2) \rightarrow w_i^{F_1}$;

Flexural$_2$ (Flex$_2$): $w_1^{SS}(x_1, x_2); w_2^{AA}(x_1, x_2); w_3^{AS}(x_1, x_2) \rightarrow w_i^{F_2}$ and

Shear (Sh): $w_1^{SA}(x_1, x_2); w_2^{AS}(x_1, x_2); w_3^{AA}(x_1, x_2) \rightarrow w_i^{Sh}$, where $S(A)$ means the parity of a function with respect to the inversion of the corresponding variable: $f(x_1, x_2) = f(-x_1, x_2) = f(x_1, -x_2) \rightarrow f^{SS}(x_1, x_2)$; $f(x_1, x_2) = -f(-x_1, x_2) = -f(x_1, -x_2) \rightarrow f^{AA}(x_1, x_2)$ etc.

The energy spectra of dilatational phonons in a homogeneous Si nanowire with the lateral cross-section 37 ML × 37 ML (1 ML = $a/4$) and Si/Ge segmented nanowire with the same cross-section and 8 atomic layers in the superlattice period (6 silicon atomic layers and 2 germanium atomic layers), calculated in the framework of FCC model, are shown in Fig. 3. The total number of phonon branches of dilatational polarization is equal to 280 for a Si NW #1 and 1120 for a Si/Ge SNW #1. In Fig. 3 the phonon branches $\hbar\omega_s(q_z)$ with quantum numbers s = 0, 1, …, 4, 10, 20, 30,…, 100, 200, 300, …, 1100 are shown. The dashed line in Fig. 3 (b) shows the maximal phonon energy in a homogeneous Ge nanowire. The maximal phonon frequency of silicon is higher than the maximal frequency of germanium, therefore high-frequency Si-like phonon modes in the Si/Ge SMW #1 are "trapped" in the Si segments and do not spread out in the Ge segments of the superlattice. These modes will not participate in the processes of heat transfer, i.e. Si/Ge SMWs act as a *phonon filter* removing many phonon modes from thermal transport [27]. Fig. 3 implies that the velocities of phonon modes with $\hbar\omega$ >7 meV in the Si nanowire are not equal to zero, whereas in a Si/Ge SNW #1, these modes are dispersionless.

<Figure 3>

The similar phonon trapping effect was also reported for cross-section modulated Si and Si/Ge MSNWs [28-29]. The phonon branches $\hbar\omega_s(q_z)$ with $s$ = 1,2,…,20, 35, 50, 65,…, 1515, 1530 of cross-section modulated Si MSNW#1 are shown in Fig. 4. As follows from Fig. 4 a great number of phonon modes in the MSNW#1 with energy $\hbar\omega$ > 5 meV are dispersionless and possess group velocities close to zero due to the trapping into the MSNW segments.

<Figure 4>

The trapping effect is illustrated in Fig. 5: the average squared displacements of atoms

$$|U(z;s,q_z)|^2 = \begin{cases} \int_{-d_{x,1}/2}^{d_{x,1}/2} \int_{-d_{y,1}/2}^{d_{y,1}/2} |\vec{w}_s(x,y,z;q_z)|^2 \, dxdy, & \text{if } 0 \leq z \leq l_1 \\ \int_{-d_{x,2}/2}^{d_{x,2}/2} \int_{-d_{y,2}/2}^{d_{y,2}/2} |\vec{w}_s(x,y,z;q_z)|^2 \, dxdy, & \text{if } l_1 < z \leq l_2 \end{cases} \quad (9)$$

in the mode ($s$=8, $q_{z}$=0.4$q_{z,\max}$) (dashed line) are relatively large in the wide segments of the MSNW and almost vanishing in the narrow segments. Therefore this mode is trapped into the wide segments of the MSNW. For comparison, the average squared displacements of atoms in a propagating phonon mode ($s$=992, $q_{z}$=0.2$q_{z,\max}$) (solid line), which are equally large in both the wide and narrow MSNW segments, is also shown.

<Figure 5>

The effect of the phonon deceleration in SNWs and MSNWs is illustrated in Figs. 6-8, where we show the average phonon group velocity

$$\langle v \rangle(\omega) = g(\omega) / \sum_{s(\omega)} (d\omega_s / dq_z)^{-1} \qquad (10)$$

as a function of the phonon energy for Si NW#2 and Si MSNW#1 (Fig. 6); Si NW#2, Si MSNW#2 and Si/Ge MSNW#1 (Fig. 7); Si NW#1, Ge NW and Si/Ge SNW#1 (Fig. 8). The average phonon group velocity in SNWs and MSNWs is smaller than that in the Si NWs for all phonon energies. As a result, the phonon modes in SNWs and MSNWs carry less heat than those in the NW. The drop in the phonon group velocities in SNWs/MSNWs in comparison with NWs is explained by the trapping effect: the trapped phonon modes represent standing waves existing only in the segments of SNWs/MSNWs.

<Figure 6>
<Figure 7>
<Figure 8>

**III. PHONON-ENGINEERED HEAT CONDUCTION OF SEGMENTED NANOWIRES.**

The phonon thermal flux per unit temperature gradient (referred as thermal flux hereafter) in 1D nanostructures is given by the expression [28-29]:

$$\Theta = \frac{1}{2\pi k_B T^2} \sum_{s=1,\ldots,3N} \int_0^{q_{z,\max}} \left(\hbar \omega_s(q_z) \upsilon_{z,s}(q_z)\right)^2 \tau_{tot,s}(q_z) \frac{\exp\left(\frac{\hbar \omega_s(q_z)}{k_B T}\right)}{\left(\exp\left(\frac{\hbar \omega_s(q_z)}{k_B T}\right) - 1\right)^2} dq_z . \qquad (11)$$

Here $\tau_{tot,s}$ is the total phonon relaxation time, $s$ is the number of a phonon branch, $k_B$ is the Boltzmann constant, $\hbar$ is the Planck constant and $T$ is the absolute temperature. The Eq. (11) is derived from the Boltzmann transport equation in the relaxation time approximation [7, 28-29, 48-50] taking into account 1D density of phonon states [27]. In case of MSNWs the phonon thermal conductivity is related to the phonon thermal flux as:

$$\kappa_{ph} = \frac{l_z^1 + l_z^2}{\left(d_x^1 + 2d_{shell}\right)\left(d_y^1 + 2d_{shell}\right) l_z^1 + \left(d_x^2 + 2d_{shell}\right)\left(d_y^2 + 2d_{shell}\right) l_z^2} \Theta . \qquad (12)$$

In case of a homogeneous MSNW $d_{shell}=0$ and equation (12) can be rewritten as: $\kappa_{ph}=\frac{l_z^1+l_z^2}{d_x^1 d_y^1 l_z^1+d_x^2 d_y^2 l_z^2}\Theta$. Finally, in case of a homogeneous NW or SNW $d_{shell}=0$, $d_x^1=d_x^2=d_x$ and $d_y^1=d_y^2=d_y$, thus equation (12) reduces to: $\kappa_{ph}=\frac{\Theta}{d_x d_y}$.

In Si and Ge nanowires there are two basic mechanisms of phonon scattering: three-phonon Umklapp scattering and boundary scattering [3, 37, 38, 48, 50-54]. According to the Matthiessen's rule, the total phonon relaxation time is given by: $1/\tau_{tot,s}(q_z)=1/\tau_{B,s}(q_z)+1/\tau_{U,s}(q_z)$. Here, $\tau_{B,s}$ is the phonon relaxation time for the boundary scattering and $\tau_{U,s}$ is the phonon relaxation time for the Umklapp scattering. The relaxation time of phonons in boundary scattering can be calculated using the following equations [28]:

$$\frac{1}{\tau_{B,s}(q_z)}=\frac{1-p}{1+p}\frac{|\upsilon_{z,s}(q_z)|}{2}\left(\frac{1}{d_x}+\frac{1}{d_y}\right) \qquad (13)$$

in the case of a NW or SNW,

$$\frac{1}{\tau_{B,s}(q_z)}=\frac{1-p}{1+p}\frac{|\upsilon_{z,s}(q_z)|}{2}\left(\left(\frac{1}{d_x^1}+\frac{1}{d_y^1}\right)\int_{-d_x^1/2}^{d_x^1/2}\int_{-d_y^1/2}^{d_y^1/2}\int_0^{l_z^1}|\vec{w}_s(x,y,z;q_z)|^2 dxdydz + \right.$$
$$\left. +\left(\frac{1}{d_x^2}+\frac{1}{d_y^2}\right)\int_{-d_x^2/2}^{d_x^2/2}\int_{-d_y^2/2}^{d_y^2/2}\int_{l_z^1+a/4}^{l_z^1+l_z^2}|\vec{w}_s(x,y,z;q_z)|^2 dxdydz\right) \qquad (14)$$

in the case of a MSNW, and

$$\frac{1}{\tau_{B,s}(q_z)}=\sum_{i=1}^{2}\frac{1}{\tau_{B,s}^i(q_z)},$$

$$\frac{1}{\tau_{B,s}^i(q_z)}=\begin{cases}\frac{1-p}{1+p}\frac{|\upsilon_{z,s}(q_z)|}{2}\left\{\xi_{core,s}^i(q_z)\left(\frac{1}{d_x^i}+\frac{1}{d_y^i}\right)+\right.\\ \left.+\xi_{shell,s}^i(q_z)\left(\frac{1}{d_x^i+2d_{shell}}+\frac{1}{d_y^i+2d_{shell}}\right)\right\}, \text{ if } \xi_{core,s}^i(q_z)/\xi_{shell,s}^i(q_z)\geq\delta,\\ \frac{1-p}{1+p}|\upsilon_{z,s}(q_z)|\frac{\xi_{shell,s}^i(q_z)}{d_{shell}}, \xi_{core,s}^i(q_z)/\xi_{shell,s}^i(q_z)<\delta,\end{cases} \qquad (15)$$

in the case of a core/shell MSNW, where

$$\xi^i_{core,s}(q_z) = \int_{-d^i_x/2}^{d^i_x/2} \int_{-d^i_y/2}^{d^i_y/2} \int_{(i-1)l^1_z}^{l^1_z+(i-1)l^2_z} |\vec{w}_s(x,y,z;q_z)|^2 \, dxdydz,$$

$$\xi^i_{shell,s}(q_z) = \int_{-(d^i_x+2d_{shell})/2}^{(d^i_x+2d_{shell})/2} \int_{-(d^i_y+2d_{shell})/2}^{(d^i_y+2d_{shell})/2} \int_{(i-1)l^1_z}^{l^1_z+(i-1)l^2_z} |\vec{w}_s(x,y,z;q_z)|^2 \, dxdydz - \xi^i_{core,s}(q_z).$$

(16)

In equations (13-15), $\upsilon_{z,s}(q_z) = d\omega_s(q_z)/dq_z$ is the phonon group velocity along the nanowire axis, $p$ is the specularity parameter of the boundary scattering. Equations (13-15) provide an extension of the standard formula for the rough edge scattering [3] to the case of a rectangular NW, SNW or MSNW. In equation (14) it was taken into account that a part of the phonon wave corresponding to the mode $(s,q_z)$, concentrated in the MSNW segment $d^1_x \times d^1_y \times l^1_z$, scatters on the boundaries of this segment, while the rest of this wave scatters on the boundaries of the segment $d^2_x \times d^2_y \times l^2_z$. In equation (15) the parameter $\delta$ was introduced in order to classify different phonon modes: core-like, shell-like and propagating. The quantities $\xi^i_{core,s}(q_z)$ and $\xi^i_{shell,s}(q_z)$ show the relative portion of the phonon mode $(s,q_z)$, concentrated in the core or shell of the $i$th MSNW segment, correspondingly. For core/shell MSNWs it is taken into account that the core-like and propagating modes partially scatter at core/shell interfaces and outer boundaries while shell-like modes with $\xi^i_{core,s}(q_z)/\xi^i_{shell,s}(q_z) < \delta$ scatter only at core/shell interfaces. In this sense, the parameter $\delta$ represents a threshold value for a ratio between the integrated phonon amplitudes concentrated in the core $\xi^i_{core,s}(q_z)$ and in the shell $\xi^i_{shell,s}(q_z)$ of $i$th MSNW segment. For example $\delta = 0.1$ means that in the shell-like modes more than 90% of lattice vibrations from the $i$th MSNW segment occur in the shell material, while core region is depleted of phonons. A similar effect of the phonon depletion was theoretically described by Pokatilov et al. in Refs. [55-56] for the acoustically-mismatched planar heterostructures, where part of the phonons is pushed out into the acoustically softer layers.

The relaxation time of phonons in Umklapp scattering can be calculated using the following formula [28-29, 51]:

$$\frac{1}{\tau_{U,s}(q_z)} = B_s(q_z)(\omega_s(q_z))^2 T \exp(-C_s(q_z)/T) \quad (17)$$

The mode-dependent parameters $B_s$ and $C_s$ in the expressions for the Umklapp scattering in core/shell MSNWs are averaged for the values of the Umklapp scattering parameters in bulk counterparts of the core and shell materials so that

$B_s(q_z) = (\xi^1_{core,s}(q_z) + \xi^2_{core,s}(q_z))B_{core} + (\xi^1_{shell,s}(q_z) + \xi^2_{shell,s}(q_z))B_{shell}$ and

$C_s(q_z) = (\xi^1_{core,s}(q_z) + \xi^2_{core,s}(q_z))C_{core} + (\xi^1_{shell,s}(q_z) + \xi^2_{shell,s}(q_z))C_{shell}$. The values $B_{core}$, $B_{shell}$, $C_{core}$,

$C_{shell}$ can be determined by comparing the calculated thermal conductivity of corresponding bulk material with experimental data. For Si and Ge there were found the following numerical values [28, 57]: $B_{Si} = 1.88 \times 10^{-19}$ s/K, $C_{Si}$ = 137.39 K, $B_{Ge} = 3.53 \times 10^{-19}$ s/K, $C_{Ge}$ = 57.6 K.

In Figure 9, the lattice thermal conductivity of Si NWs and Si MSNWs are plotted as a function of temperature for Si NW #2, as well as for Si MSNWs #2, #3, #4 and #5. The results are presented for a reasonable specularity parameter $p$ = 0.85, which was found in Ref. [38] from a comparison between theoretical and experimental data for a Si film of 20 nm thickness.

<Figure 9>

A significant redistribution of the phonon energy spectra and a reduction of the average phonon group velocities in MSNWs strongly suppress their lattice thermal conductivity in comparison with the NW. At room temperature, the ratio between the thermal conductivities in NW and MSNWs ranges from a factor of 5 to 13 depending on the cross-section $S_2 = d_x^2 \times d_y^2$. However, in order to compare more correctly the abilities of MSNWs and NWs to conduct heat one should compare thermal fluxes $\Theta$ rather than thermal conductivities $\kappa_{ph}$ since latter depend explicitly on the dimensions of the nanowires.
In Figure 10(a) we show the thermal flux for Si NW #2 (upper dashed line) and Si MSNWs #3, #4 and #6 for $p$ = 0.85 as a function of temperature.

<Figure 10>

The maxima on the thermal flux curves are determined by the interplay between the three-phonon Umklapp and the phonon-boundary scattering. At low temperatures, the boundary scattering dominates; the thermal flux increases with temperature due to the population of high-energy phonon modes and approaches the maximum value when $\tau_U \sim \tau_B$. A further temperature rise leads to an enhancement of the Umklapp scattering and diminution of the thermal flux. An increase of the cross-section of the MSNW wide segments attenuates the phonon-boundary scattering, and the maximum of the thermal flux curves shifts to lower temperatures: from $T$ = 190 K for Si NW #2 to $T$ = 100 K for Si MSNW #6. Therefore, at low temperatures ($T$<120 K) the thermal flux reduction is stronger in MSNWs with the smaller cross-sections. A large number of high energy phonon modes in MSNWs are trapped in the wider segments and possess group velocity close to zero. The population of these modes with temperature rise almost does not increase the thermal flux. Thus, at medium and high temperatures the Umklapp-limited thermal flux in MSNWs reduces stronger than that in the NWs without modulation. The ratio of the thermal fluxes in NW and MSNW $\eta = \Theta(\text{Si NW})/\Theta(\text{Si MSNW})$ increases with temperature, and reaches the values of 3.5 to 4.5 depending on the MSNW cross-section (see Figure 10(b)). At temperatures above 150 K the increase of the MSNW cross-section makes the reduction of the thermal flux stronger due to the corresponding rise of the number of the trapped high-energy phonon modes, which do not carry heat in MSNWs. This is distinct from the case of NWs without cross-section modulation. The calculations also showed that the strong modification of the phonon energy spectra and phonon group velocities in MSNWs in comparison with NWs also increases the Umklapp phonon scattering, which is an additional reason for the thermal flux reduction in MSNWs.

An important quantity, which determines the thermal conductivity and thermal flux, is the mode-dependent phonon mean free path (MFP) $\Lambda_s(q_z)$ [58]. Following the Matthiessen's rule the total phonon MFP $\Lambda_s(q_z)$ is given by

$$1/\Lambda_s(q_z) = \sum_{r=B,U} 1/\Lambda_{r,s}(q_z), \qquad (18)$$

where $\Lambda_{r,s}(q_z) = \tau_{r,s}(q_z) \cdot \upsilon_{z,s}(q_z)$ and $r=B$ stands for phonon-boundary scattering, while $r=U$ for three-phonon Umklapp scattering. The dependence of the average phonon MFP $\langle \Lambda \rangle(\omega) = g(\omega)/\sum_{s(\omega)}(1/\Lambda_s)$ on the phonon energy is presented in Figure 11 for the Si NW #2 (dashed line) and Si MSNWs #7 (solid line) and #8 (dotted line).

<Figure 11>

The Umklapp-limited phonon MFPs $\Lambda_{U,s}(q_z)$ in MSNWs are smaller than those in NWs due to both reduced phonon group velocity and stronger phonon scattering. The boundary-limited MFPs $\Lambda_{B,s}(q_z)$ are larger in MSNWs due to the larger average cross-section of MSNW in comparison with that in NW (see equations (14, 15)). As a result, at small energies when the Umklapp scattering is weaker than the boundary scattering $\langle \Lambda \rangle^{MNW} > \langle \Lambda \rangle^{NW}$, while for $\hbar\omega > 5$ meV $\langle \Lambda \rangle^{MNW} \ll \langle \Lambda \rangle^{NW}$. The augmentation of $l_z^2$ decreases $\langle \Lambda \rangle^{MNW}$ for almost all phonon energies. The energy-averaged phonon MFP calculated from Figure 11 constitutes ~ 9.25 nm for the Si NW #2, ~ 8.4 nm for the Si MSNW #7 and ~ 6.9 nm for the Si MSNW #8. The increase of the MSNW average cross-section at fixed $l_z^1$ and $l_z^2$ attenuates boundary scattering and increases the thermal flux. In Figure 12 is shown the dependence of the ratio $\eta$ of the thermal fluxes in Si NW #2 and Si MSNW #2 on temperature for different values of the specularity parameter $p = 0.0, 0,3, 0.6$, and $0.9$.

<Figure 12>

For the interpretation of the data in Figure 12, authors of Refs. [28] were calculated separately the thermal flux $\Theta_B$ carried out by the long-wavelength phonon modes $(s,q_z)$, which are mainly scattered at the boundaries and described by the inequality $\tau_U(s,q_z) \geq \tau_B(s,q_z)$, and the thermal flux $\Theta_U$ carried out by the rest of the phonons i.e. thermal flux mainly limited by Umklapp scattering processes. The total thermal flux is thus: $\Theta = \Theta_B + \Theta_U$. The calculations showed that for all values of $p$ under consideration, the room temperature thermal flux $\Theta_B$ is by a factor of ~5 lower in MSNW than that in NW due to the phonon trapping. An increase of $p$ decreases $\Theta_B$ and strongly enhances $\Theta_U$ in NW due to attenuation of the boundary scattering of the high-energy phonon modes. These phonons in MSNW do not participate in the heat transfer because of their localization in the wider segments. For this reason, the ratio between thermal fluxes in the NW and the MSNW appreciably depends on $p$: for $p = 0$ $\Theta_U(NW)/\Theta_U(MSNW) \sim 1$, while for $p = 0.9$ $\Theta_U(NW)/\Theta_U(MSNW) \sim 3$. As a result, the flux ratio increases with increasing $p$ in a wide range of temperatures from 100 K to 400 K. Considering that the room temperature thermal conductivity of the rough Si NWs [59] is already by a factor of 100 lower than the corresponding bulk value, the obtained results suggest that the cross-section modulation of the rough Si NWs will allow for an additional decrease of the thermal conductivity by a factor of 2 to 2.5 with a subsequent increase of the thermoelectric efficiency *ZT*.

The dependence of the ratio $\eta$ of the thermal fluxes in Si NW #2 and Si MSNW 14 ML x 14 ML x $N_z$ / 22 ML x 22 ML x $N_z$ on $N_z$ for the temperatures $T = 100$ K, $T = 200$ K, $T = 300$ K and $T = 400$ K and $p = 0.85$ is presented in Figure 13. The calculated points for $N_z = 2,4,6,…,18$ are joined by the smooth curves as guides for an eye.

<Figure 13>

The overall trend of these curves is determined by the interplay of two effects: (i) phonon trapping, which suppresses the thermal flux and (ii) augmentation of the MSNW average cross-section, which enhances the flux due to the emergence of additional phonon modes for heat propagation and attenuation of the phonon-boundary scattering. In Si MSNW with the ultra-narrow segments $N_z = 2$ ML, the trapping of phonon modes is weak and the thermal flux is larger than that in Si NW ($\eta < 1$). The rise of $N_z$ enhances the trapping, and for all temperatures under consideration the flux ratios rapidly increase with $N_z$ rising up to the values 8 ML to 12 ML, and reach their maximum values at around $N_z = 16$ ML to 18 ML. It is expected that a subsequent rise of $N_z$ should decrease $\eta$ due to augmentation of the MSNW average cross-section.

In Figure 14 is presented the temperature dependence of phonon thermal conductivity for Si NW #2, Si MSNW #2 and Si MSNWs covered by Ge shell of different thickness $d_{Ge}$.

<Figure 14>

Increasing the thickness of Ge shell from 1 ML to 7 ML leads to a decrease of thermal conductivity of Si/Ge MSNW by a factor of 2.9 – 4.8 in comparison with that in Si MSNW without Ge, and by a factor of 13 – 38 in comparison with that in Si NW. The reduction in κ of Si/Ge MSNWs is substantially stronger than that reported for core/shell nanowires without cross-sectional modulation [15, 38, 60-63]. In the core/shell nanowires without modulation, the κ decrease is mainly due to phonon hybridization, which results in changes in the phonon DOS and group velocity. In the core/shell MSNWs, reduction of thermal conductivity is reinforced due to localization of a part of phonon modes in wider MSNW segments. The localization completely removes such phonons from the heat transport.

The dependence of the ratio of thermal fluxes at room temperature for MSNWs and Si NW on $d_{Ge}$ is presented in Figure 15 for $N_z = 4, 6, 8, 12$ ML. Two points for larger $N_z = 20$ ML and 28 ML at $d_{Ge} = 4$ ML are also shown. All curves have a maximum between $d_{Ge} = 3$ ML and $d_{Ge} = 6$ ML. The increase in $N_z$ leads to a shift of the maximum to lower values of $d_{Ge}$.

<Figure 15>

In order to explain the dependence presented in Figure 15 it was analyzed the frequency-specific thermal flux $\phi(\omega)$, i.e. the integrand in the equation $\Theta = \int_0^{\omega_{max}} \phi(\omega) d\omega$. The results show that in Si MSNWs without Ge shell, thermal flux is strongly suppressed in comparison with Si NW due to redistribution of phonon energy spectra leading to the reduction of the phonon group velocities and localization of a part of phonon modes in nanowire segments. The influence of Ge shell on the frequency-specific thermal flux of MSNWs is determined by two opposite effects: (i) it reinforces the decrease of the thermal flux in Si/Ge MSNWs as compared with Si MSNWs owing to a stronger decrease of the phonon velocities and stronger phonon localizations; (ii) it increases the thermal flux due to appearance of additional channels for heat transfer through the Ge shell and weakening of the phonon-boundary scattering of propagating and Ge-like phonons. An interplay between two effects explains the non-monotonic dependence of the thermal flux ratio on $d_{Ge}$ shown in Figure 15. The difference between thermal fluxes in MSNWs and NWs becomes larger with growing $N_z$ and reaches maximum value of ~10 at $N_z$ ~ 28 ML – 32 ML. For $N_z$>32 ML, thermal flux ratio starts to decrease due to redistribution of the phonon energy spectra and heat conduction through Ge shell.

For ideally smooth interfaces when all phonon scattering events are specular $p=1$. The value of $p=0.85$, used in the calculations, corresponds to smooth NW surfaces with root mean square height of interface roughness $\Delta \sim 1 ML$. The $\Delta$ was estimated by averaging the mode-dependent specular parameter $\tilde{p}(q,\Delta) = \exp(-2q^2\Delta^2)$ [3, 12, 64, 65] over all $q$: $p = \int_0^{q_{z,\max}} \tilde{p}(q)dq / q_{z,\max}$. The small roughness of NWs and MSNWs interfaces is beneficial for both maintaining high electron mobility [66, 67] and for suppression of the phonon heat conduction in MSNWs [56]. The increase of $p$ leads to faster growth of the thermal flux in Si NW than in Si MSNWs due to exclusion of high energy phonons in MSNW from heat transfer.

In Fig. 16 is plotted the temperature dependence of the phonon thermal conductivity of Si/Ge SNWs with cross-section 37 ML x 37 ML and different lengths of the Si and Ge segments along the wire axis. The results for Si NW #1 and Ge NW with the same cross-section are also presented for comparison. In the temperature range 150 K - 300 K the thermal conductivity in the Si/Ge SNW is 5 - 6 times lower than that in the Ge nanowire with the same cross section, and 9 - 11 times lower than that in the Si nanowire.

<Figure 16>

When the number of atomic layers of Si per period increases from 8 to 12, the properties of the Si/Ge SNW reveal a slight trend towards those of the Si nanowire. Therefore the phonon thermal conductivity of the Si/Ge SNW, containing an equal number of atomic layers of Si and Ge, is lower than that of SNW containing different numbers of atomic layers per period. In Ref. [68] it was theoretically shown that the thermal conductivity of SNWs, consisting of different isotopes of silicon is by a factor of 2 smaller than in a Si nanowire. Our results demonstrate an even greater drop in the thermal conductivity in SNW composed of segments from acoustically-mismatched materials due to a stronger localization of phonon modes in the superlattice segments and a stronger decrease of the phonon group velocities. Our findings for the Si/Ge SNWs are in a qualitative agreement with the reduction of the thermal conductivity below the alloy limit predicted for circular Si/Ge SNWs with diameters less than 15 nm [69]. The thermal conductivity down to the sub-1 $Wm^{-1}K^{-1}$ range was achieved in multilayered Ge/Si dot arrays [70]. A similar effect of a strong decrease of the thermal conductivity is demonstrated theoretically in Ref. [71] for Si/Ge 3D-SNWs. The authors of Ref. [71] explain the thermal conductivity reduction by a significant decrease of the phonon group velocities and incoherent scattering of phonons.

## IV. CONCLUSIONS

We reviewed recent theoretical results on phonon heat conduction in Si and Si/Ge segmented nanowires in the framework of the face-centered cubic cell and Born-von Karman models of lattice dynamics. Trapping of the phonon modes in the nanowire segments, leading to redistribution of the phonon energy spectra, decrease of the phonon group velocities and suppression of the phonon heat conduction are discussed in details. The room temperature heat flux in segmented nanowires can be suppressed by almost three orders-of-magnitude in comparison with that in bulk Si and by an order-of-magnitude in comparison with that in generic Si nanowires. We argue that geometry modulation and acoustic mismatch are highly efficient instruments in engineering phonons in semiconductor segmented nanowires for their thermoelectric and thermal insulator applications.

**Acknowledgement**

DLN and AIC acknowledge the financial support from the Republic of Moldova through the projects 15.817.02.29F and 14.820.18.02.012 STCU.A/5937 and from the Science and Technology Center in Ukraine (STCU, project #5937). The work at the University of California – Riverside was supported by the National Science Foundation.

| Nanostructure | Dimensions | Notation in the present work |
|---|---|---|
| Si NW | 37 ML × 37 ML | Si NW #1 |
| Si NW | 14 ML × 14 ML | Si NW #2 |
| Ge NW | 37 ML × 37 ML | Ge NW |
| Si/Ge SNW | 37 ML × 37 ML × 6 ML / 37 ML × 37 ML × 2 ML | Si/Ge SNW#1 |
| Si MSNW | 14 ML × 14 ML × 6 ML / 22 ML × 22 ML × 6 ML | Si MSNW #1 |
| Si MSNW | 14 ML × 14 ML × 8 ML / 22 ML × 22 ML × 8 ML | Si MSNW #2 |
| Si MSNW | 14 ML × 14 ML × 8 ML / 18 ML × 18 ML × 8 ML | Si MSNW #3 |
| Si MSNW | 14 ML × 14 ML × 8 ML / 26 ML × 26 ML × 8 ML | Si MSNW #4 |
| Si MSNW | 14 ML × 14 ML × 8 ML / 30 ML × 30 ML × 8 ML | Si MSNW #5 |
| Si MSNW | 14 ML × 14 ML × 8 ML / 34 ML × 34 ML × 8 ML | Si MSNW #6 |
| Si MSNW | 14 ML × 14 ML × 4 ML / 22 ML × 22 ML × 4 ML | Si MSNW #7 |
| Si MSNW | 14 ML × 14 ML × 12 ML / 22 ML × 22 ML × 12 ML | Si MSNW #8 |
| Si/Ge MSNW | 14 ML × 14 ML × 8 ML / 22 ML × 22 ML × 8 ML – $d_{Ge}$=4 ML | Si/Ge MSNW #1 |

**Table 1 of 2.** Notations and dimensions of NWs, SNWs and MSNWs under consideration.

| Components of vectors $\vec{h}_n^I$ for the selected "white" atom $\vec{r}_k$ | Components of vectors $\vec{h}_n^I$ for the selected "dark" atom $\vec{r}_k$ | Components of vectors $\vec{h}_n^{II}$ for the selected "white" (or "dark") atom $\vec{r}_k$ | Set of five force constants of Si and Ge used for calculation (N/m) |
|---|---|---|---|
| $a/4(1,1,1)$; $a/4(1,-1,-1)$; $a/4(-1,1-1)$; $a/4(-1,-1,1)$ | $a/4(-1,-1,-1)$; $a/4(-1,1,1)$; $a/4(1,-11)$; $a/4(1, 1,-1)$ | $a/2(1,1,0)$; $a/2(-1,-1,0)$; $a/2(1,0,1)$; $a/2(-1,0,-1)$; $a/2(0,1,1)$; $a/2(0,-1,-1)$; $a/2(-1,1,0)$; $a/2(1,-1,0)$; $a/2(-1,0,1)$; $a/2(1,0,-1)$; $a/2(0,-1,1)$; $a/2(0, 1,-1)$ | Silicon parameters: $\alpha = 54.85$ $\beta = 35.0$ $\mu = 3.8$ $\nu = 2.5$ $\lambda = -4.42$ Germanium parameters: $\alpha = 49.6$ $\beta = 33.0$ $\mu = 3.03$ $\nu = 3.03$ $\lambda = -3.0$ |

**Table 2 of 2**. Components of the nearest and second-nearest atoms $\vec{r}'_{k,n} = \vec{r}_k + \vec{h}_n^{I(II)}$ in the diamond-type unit cell and a set of the force constants used for Si. The table is adopted from Ref. [28] with permission from American Physical Society.

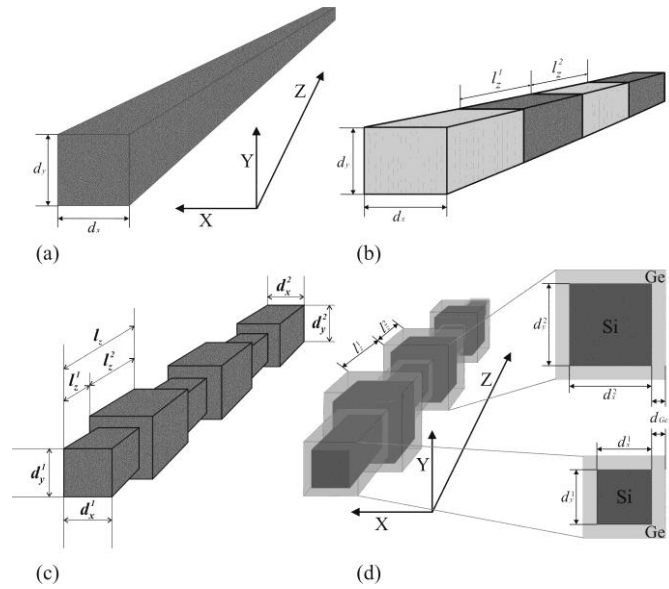

**Figure 1 of 16.** Schematic view of considered generic Si nanowire (a), Si/Ge segmented nanowire (b) and cross-section-modulated Si (c) and Si/Ge (d) nanowires. The image was adopted from Refs. [27-29] with permission from American Physical Society and American Institute of Physics.

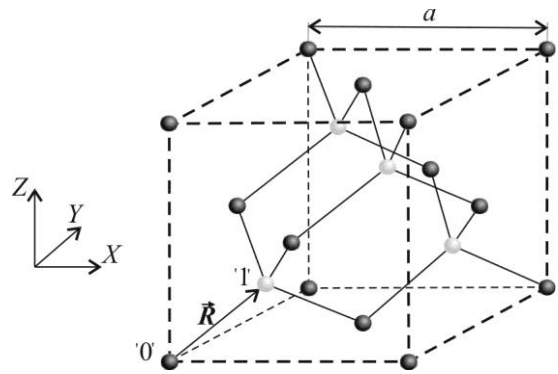

**Figure. 2 of 16**. Schematic view of silicon crystal lattice. White and black atoms show atoms from different face-centered cubic sub-lattices.

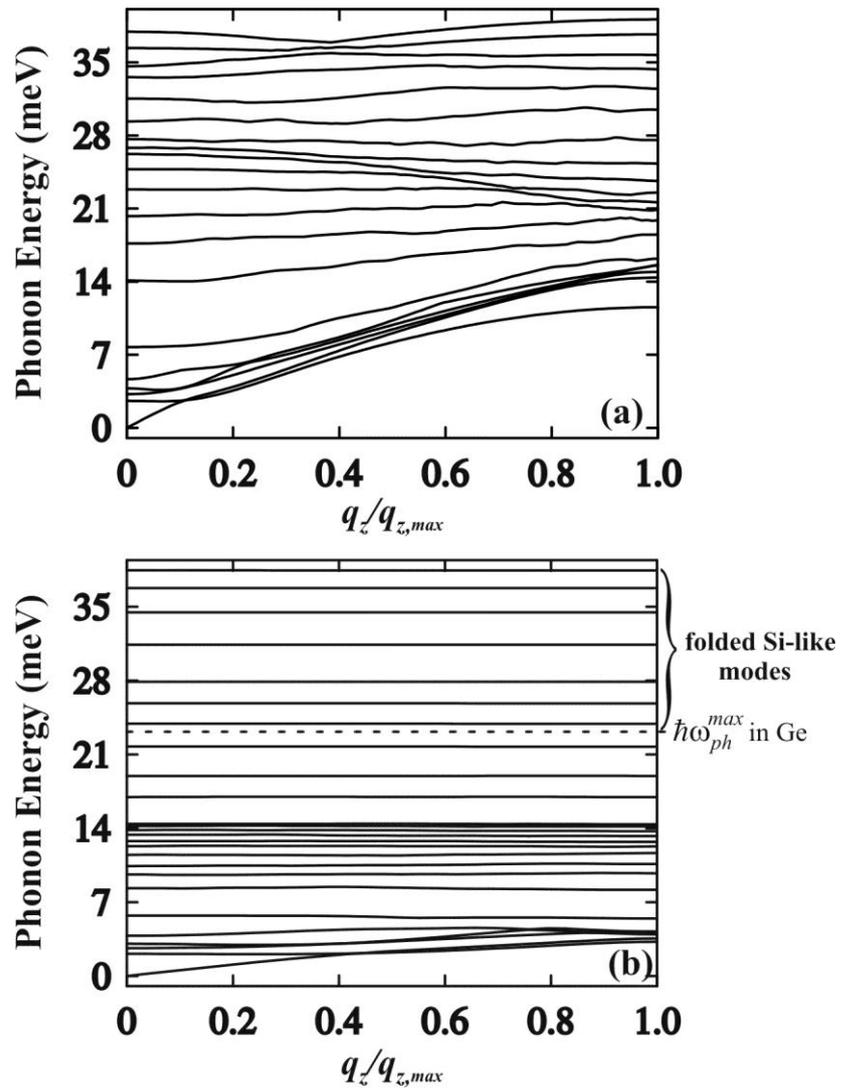

**Figure. 3 of 16.** Dilatational phonon energies as a function of the phonon wave vector $q$ in (a) a homogeneous rectangular Si nanowire with the lateral cross-section 37 ML × 37 ML nm. The phonon branches with $s$ = 0 to 4, 10, 30, 50…280 are shown; (b) a Si/Ge SNW with the same lateral cross section and 8 atomic layers per superlattice period (2 atomic layers of Ge and 6 atomic layers of Si). The phonon branches with $s$=0 to 4, 10, 30, 50…190, 200, 300,…1100, 1120 are depicted. The image is reprinted from Ref. [27] with permission from American Physical Society.

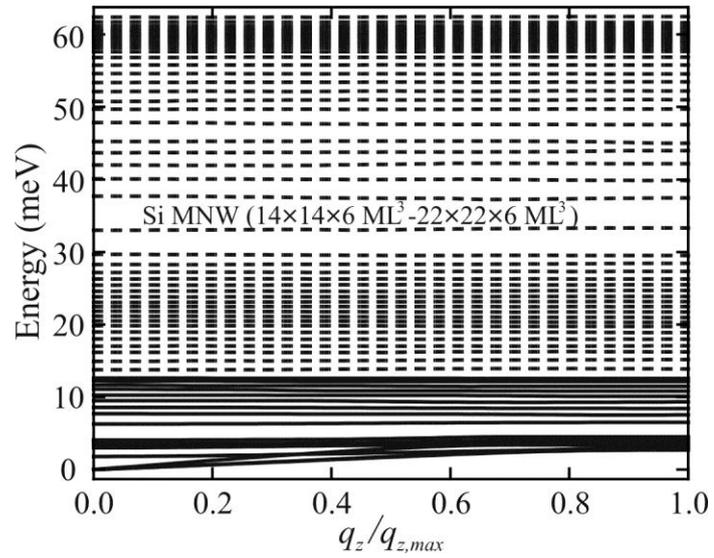

**Figure 4 of 16**. Phonon energies as a function of the phonon wave vector $q$ in Si MSNW with dimensions 14 ML × 14 ML × 6 ML/ 22 ML × 22 ML × 6 ML. The phonon branches with $s$=1 to 20, 35, 50, 65, …, 1515, 1530 are depicted. The image is reprinted from Ref. [28] with permission from American Physical Society.

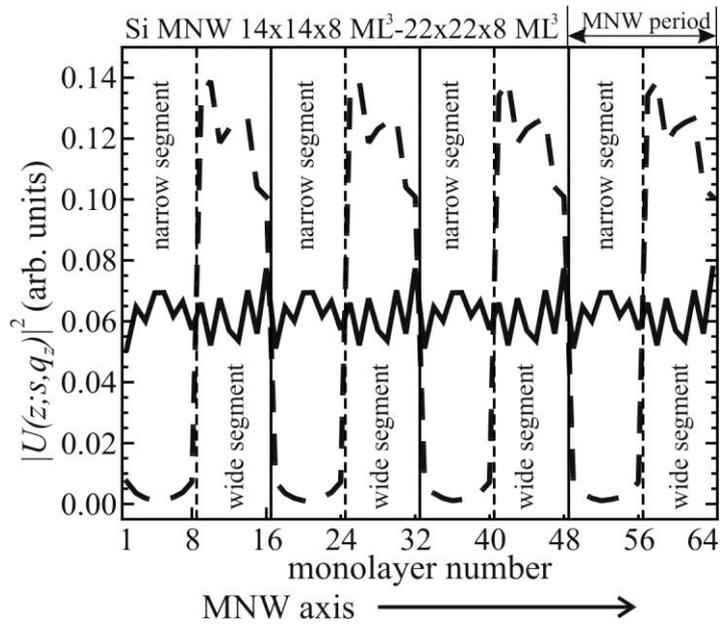

**Figure 5 of 16**. Average squared displacements $|U(z;s,q_z)|^2$ of the trapped ($s=8$, $q_z=0.4q_{z,max}$) (dashed line) and propagating ($s=992$, $q_z=0.2q_{z,max}$) (solid line) phonon modes in Si MNW with dimensions 14 ML × 14 ML × 6 ML/ 22 ML × 22 ML × 6 ML. The image is reprinted from Ref. [28] with permission from American Physical Society.

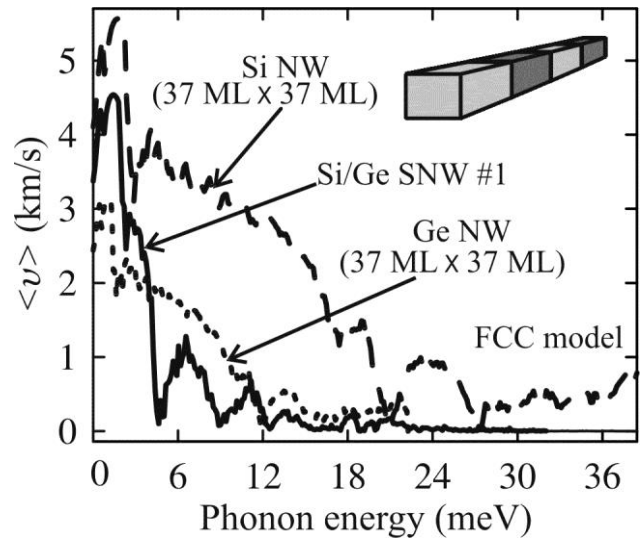

**Figure 6 of 16.** Average phonon group velocity as a function of the phonon energy in Si NW with the lateral cross-section 14 ML × 14 ML and Si MSNW with dimensions 14 ML × 14 ML × 6 ML/ 22 ML × 22 ML × 6 ML. The image is reprinted from Ref. [28] with permission from American Physical Society.

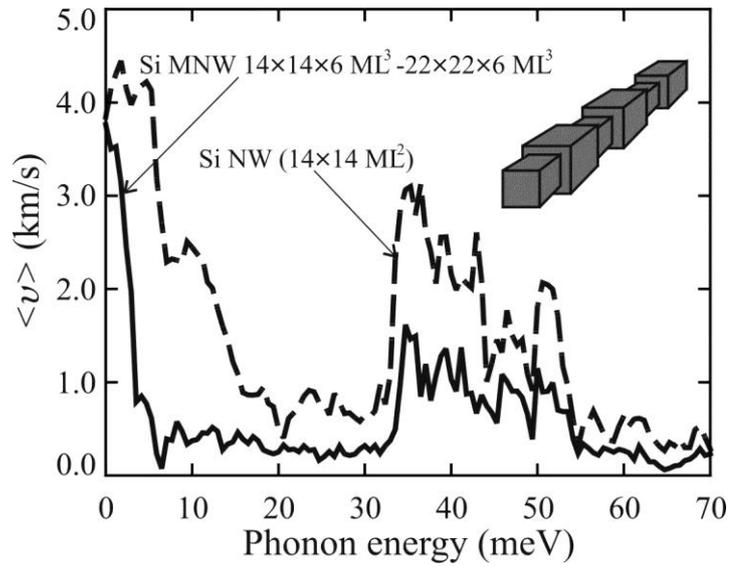

**Figure 7 of 16**. Average phonon group velocity as a function of the phonon energy shown for Si NW with the lateral cross-section area 14×14 ML, Si MSNW with dimensions 14×14×8−22×22×8 ML and Si/Ge core-shell MSNWs with 14×14×8−22×22×8 ML Si core and $d_{Ge}$ = 4 ML. The image is reprinted from Ref. [29] with permission from American Institute of Physics.

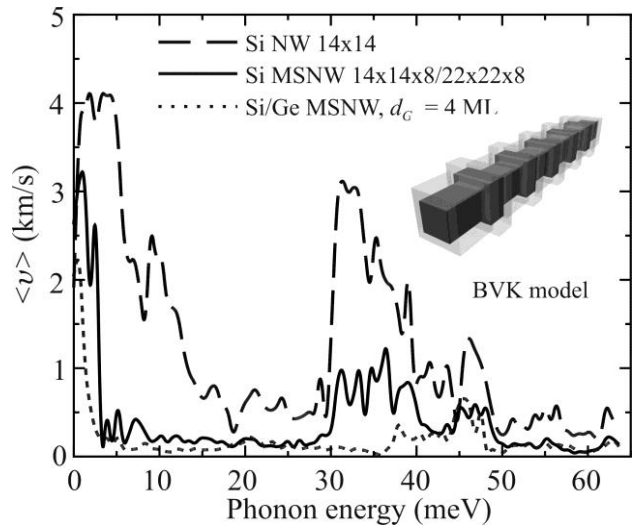

**Figure 8 of 16**. Average phonon group velocity as a function of the phonon energy in Si and Ge homogemeous nanowires with the lateral cross-section 37 ML and in Si/Ge SNW with the same lateral cross-section and 8 atomic layers per superlattice period (2 atomic layers of Ge and 6 atomic layers of Si). The image is reprinted from Ref. [27] with permission from American Physical Society.

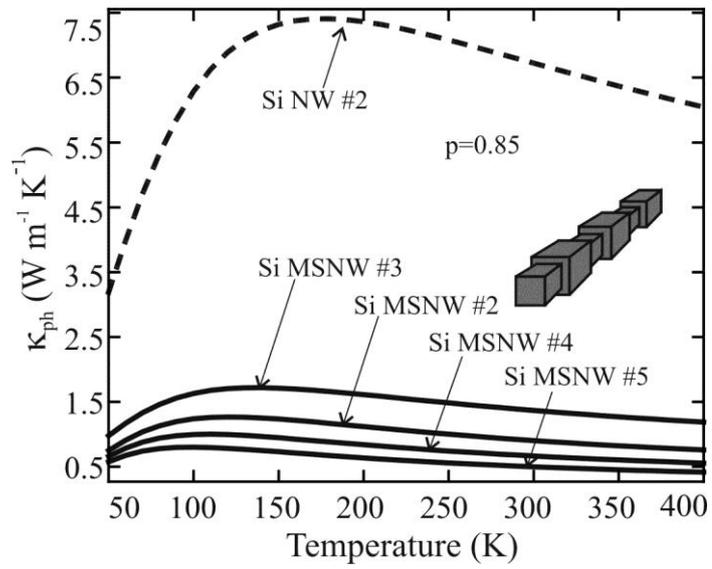

**Figure 9 of 16**. Temperature dependence of the lattice thermal conductivity in Si NW with the cross-section 14 ML × 14 ML and Si MSNW with different dimensions. The image is reprinted from Ref. [28] with permission from American Physical Society.

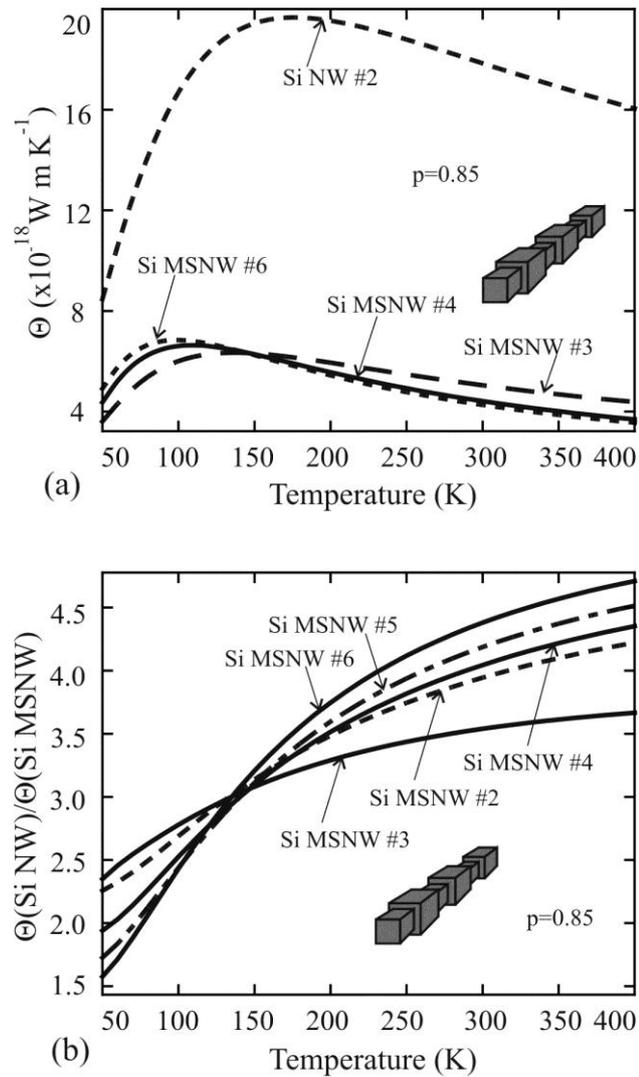

**Figure 10 of 16**. (a) Temperature dependence of the thermal flux for Si NW (dashed line) and Si MSNWs with different dimensions. (b) Temperature dependence of the ratio between thermal fluxes in Si NW and Si MSNWs. The image is reprinted from Ref. [28] with permission from American Physical Society.

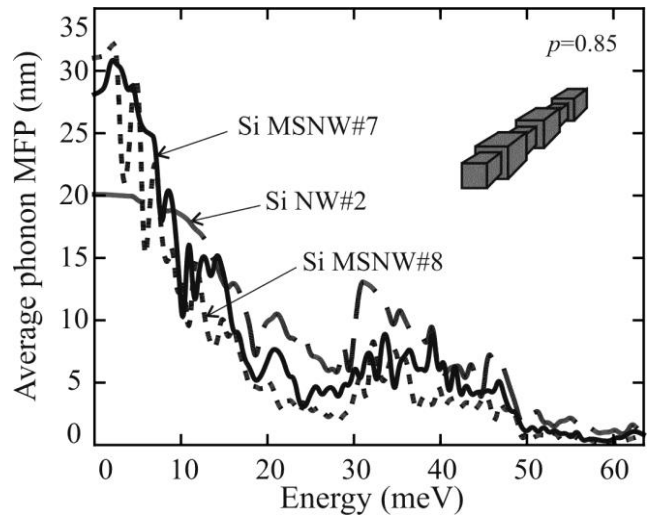

**Figure 11 of 16**. Dependence of the average phonon mean free path on the phonon energy in Si NW (dashed line), Si MSNW with $N_z$=4 ML (solid line) and Si MSNW with $N_z$=12 ML (dotted line). The image is reprinted from Ref. [28] with permission from American Physical Society.

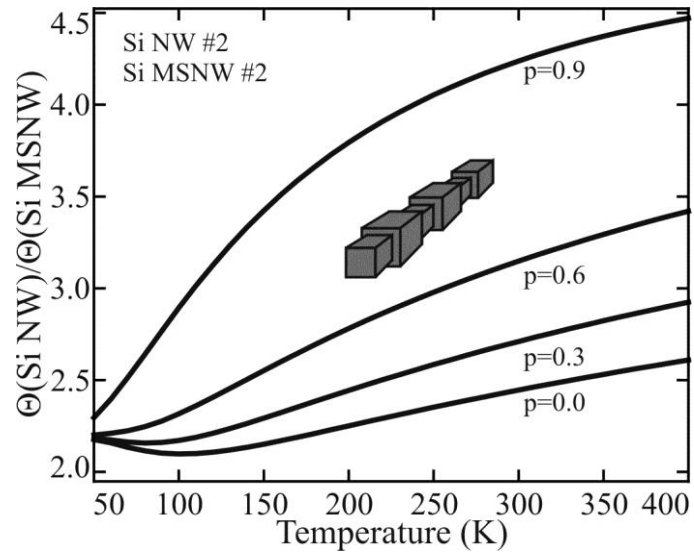

**Figure 12 of 16**. Temperature dependence of the ratio of thermal fluxes in Si NW and Si MSNWs. The results are shown for different values of the specularity parameter $p$ = 0.0, 0.3, 0.6 and 0.9. The image is reprinted from Ref. [28] with permission from American Physical Society.

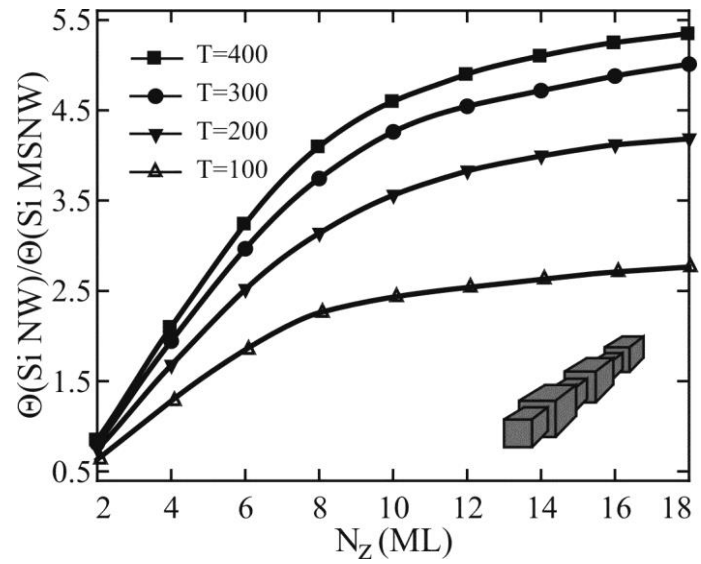

**Figure 13 of 16**. Ratio of thermal fluxes in Si NW and Si MSNWs as a function of $N_z$. The results are shown for different temperatures $T$ = 100 K, 200 K, 300 K and 400 K. The image is reprinted from Ref. [28] with permission from American Physical Society.

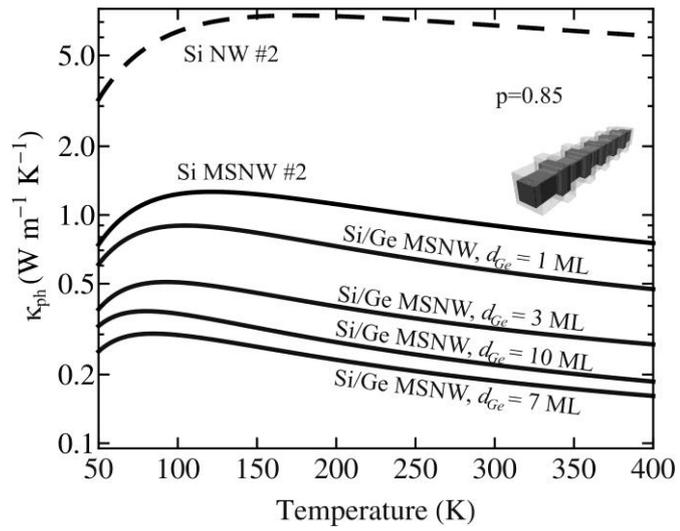

**Figure 14 of 16**. Phonon thermal conductivity as a function of the absolute temperature. Results are presented for Si NW, Si MSNW and core/shell Si/Ge MSNW with different thickness of Ge. The image is reprinted from Ref. [29] with permission from American Institute of Physics.

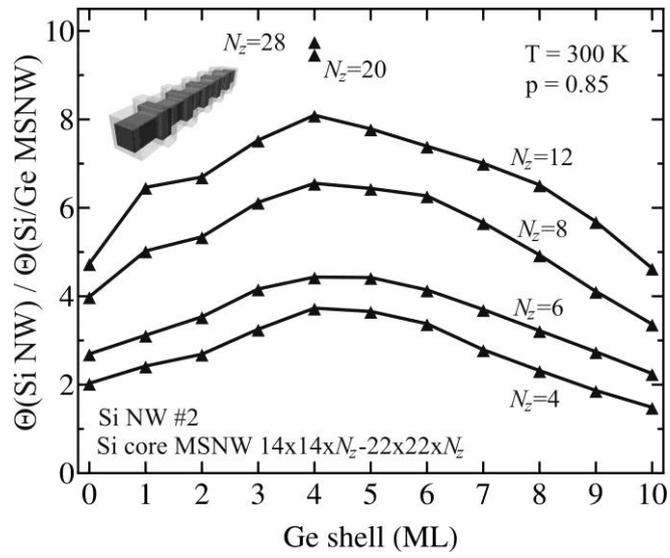

**Figure 15 of 16**. Ratio of thermal fluxes in Si NW and Si/Ge MSNWs as a function of $d_{Ge}$ for different values of $N_z$. The image is reprinted from Ref. [29] with permission from American Institute of Physics.

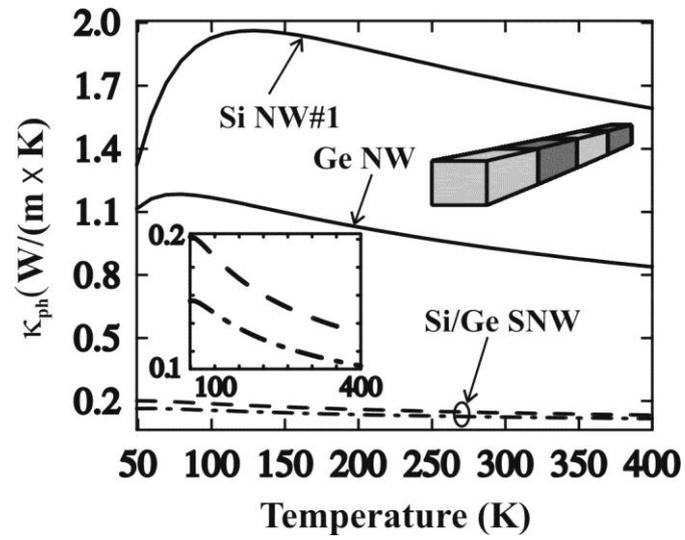

**Figure 16 of 16**. Temperature dependence of lattice thermal conductivity for Si and Ge homogeneous nanowires (solid lines) and for Si/Ge SNWs with 12 ML of Si and 4 ML of Ge (dashed line) and with 8 ML of Si and 8 ML of Ge (dash-dotted line) per period. The image is reprinted from Ref. [27] with permission from American Physical Society.